\documentstyle[aps,epsfig]{revtex}

\begin{document}

\bibliographystyle{unsrt}
\title{A Transverse Oscillation Arising From Spatial Soliton
Formation in Nonlinear Optical Cavities}
\author{Jack Boyce and Raymond Y. Chiao \\
Department of Physics, University of California, Berkeley, 
California 94720 \\
Voice: (510) 642-5620 Fax: (510) 642-5620}
\date{This version was produced on \today}
\maketitle
\begin{abstract}
A new type of transverse instability in dispersively nonlinear optical 
cavities, called the \textit{optical whistle}, is discussed.  This 
instability occurs in the mean field, soliton forming limit when the 
cavity is driven with a finite width Gaussian beam, and gives rise to 
oscillation, period doubling, and chaos.  It is also seen that 
bistability is strongly affected due to the oscillation within the 
upper transmission branch.  The phenomenon is interpreted as a 
mode mismatch in the soliton formation process and is believed to have
broad applicability.
\end{abstract}

\section*{Introduction}

Over the last 25 years the nonlinear optical resonator has been a rich 
source of interesting phenomena \cite{reinisch1994}.  These systems 
are attracting attention recently due to their potential applications 
as all-optical switches, memories \cite{firth1996}, and logic gates 
\cite{turchette1995}, at both the classical and quantum levels.  The 
resonator geometry naturally gives appreciably stronger nonlinear 
effects for a given incident beam intensity than do the 
travelling wave schemes, potentially allowing the use of faster, less 
lossy, materials with relatively weaker nonlinearities (for example, 
atomic vapors or even silica as opposed to thermal or photorefractive 
materials).  The resonator geometry also leads to a variety of 
time-dependent behaviors which must be adequately characterized, 
either to avoid undesired effects in engineered systems or to perhaps 
take advantage of them for communications or computing purposes.

The most fundamental physical process in the nonlinear resonator, 
plane wave optical bistability, was first observed experimentally in 
1975 \cite{mccall1975}.  Ikeda \cite{ikeda1979} later showed in a 
discrete return-map analysis that as the incident intensity increases 
the intracavity field intensity can oscillate and undergo a period 
doubling route to chaos.  In the 1980's a great deal of work 
concentrated on transverse structure and dynamics in nonlinear 
cavities, in order to incorporate the phenomena of self-(de)focusing 
and diffraction.  McLaughlin {\it et al} \cite{mclaughlin1985} 
demonstrated in 1985 that a finite-width Gaussian beam incident on a 
cavity with a single transverse dimension can give rise to a {\it 
transverse} oscillation in the cavity field, essentially due to a 
modulational instability.  These oscillations are of period equal to a 
multiple of the cavity round-trip time and thus correspond to the 
occupation of multiple resonator longitudinal modes.  More recently 
Haelterman and Vitrant discussed a transverse oscillation which arises 
when a nonlinear Fabry-Perot is illuminated obliquely, due to the 
lateral drift of the field within the cavity \cite{haelterman1992}.

Here we report on the discovery of a new type of oscillation within 
high-finesse nonlinear cavities driven by Gaussian beams.  This 
\textit{optical whistle} oscillation occurs as a competition between 
spatial soliton formation \cite{chiao1964}\cite{akhmanov1966} and the 
driving input, and can exhibit oscillatory or chaotic 
behavior.

\section*{General Considerations}

The dynamics of a general (dispersively) nonlinear optical cavity driven
with a finite-width beam is principally governed by four effects:
propagation, damping, diffraction, and nonlinearity.  A qualitative
understanding of cavity behavior is gained by simply comparing the
characteristic timescales of these processes, given by:
\begin{equation}
\begin{array}{rcl}
\tau_{prop}&=&L/c \\
\tau_{damp}&=&L{\cal F}/c \\
\tau_{diff}&=&w^{2}/\lambda c \\
\tau_{nl}&=&\lambda/c\Delta n
\end{array}
\label{timescales}
\end{equation}
where $L$ and $\cal F$ are the cavity length and finesse, $\lambda$ is
the optical wavelength, $w$ is a characteristic size of transverse
features in the optical beam, and $\Delta n = |n(I)-n(0)|$ is the
typical nonlinear index shift caused by the optical field of typical
intensity $I$ (the linear index is assumed to be unity).

The timescale $\tau_{nl}$ is the time required for the nonlinearity to 
induce a $2\pi$ phase shift in the light field.  For self-focusing 
nonlinearities we can also identify $\tau_{nl}$ as the typical 
formation time of transverse structure arising from the modulational 
instability \cite{benjamin1966}.  These structures form with 
characteristic transverse wavelength
$\Lambda = \lambda/\sqrt{\Delta n}$.  A feature of this size will
diffract apart in a time $\Lambda^{2}/(\lambda c)\approx\tau_{nl}$, 
and we generally conclude that these structures spontaneously form so 
that diffraction and nonlinearity are equally important.  The same 
result applies to the formation of spatial solitons, which can be 
considered modes of self-induced waveguides 
\cite{snyder1995}\cite{snyder1997}.  Soliton formation is more 
efficient when the incident beam is of size $\Lambda$, in analogy with 
mode-matching in linear optics.  The optical whistle instability 
arises from a ``mode mismatch'' between the incident beam size and the 
(self-consistent) soliton dimension $\Lambda$.

Returning to the cavity timescales of Eq. \ref{timescales}, one 
important case is when $\tau_{prop}$ is much shorter than the others.
In this \textit{mean field limit} the field changes little in
traversing the cavity once, and we can suppress the field envelope's
longitudinal dependence.  It can be
shown that in this limit only a single longitudinal
cavity mode is appreciably occupied \cite{lugiato1988}.

The optical whistle phenomenon exists under the same conditions as spatial
soliton formation:
\begin{equation}
\tau_{prop}\ll\tau_{diff}\approx\tau_{nl}\ll\tau_{damp} 
\label{conditions}
\end{equation}
For the remainder of the discussion we confine ourselves to this case.
From Eq. \ref{timescales} it follows that we are discussing a
high-finesse cavity.

\section*{The Nonlinear Cavity Equation}

We now restrict our attention to a Kerr nonlinear cavity having a single 
transverse degree of freedom $x$.  The cavity's internal field envelope 
is governed by the Lugiato-Lefever equation \cite{lugiato1987}, 
written here as
\begin{equation}
\frac{\partial{\cal E}}{\partial t} = 
\frac{i c}{2 k}\frac{\partial^2 {\cal E}}{\partial x^2} - \Gamma'{\cal E}
+i(\Delta\omega'){\cal E} + \Gamma'{\cal E}_{drive} + 
i\omega n_2 |{\cal E}|^2{\cal E}\,,
\label{dimcavityeqn}
\end{equation}
where ${\cal E}$ is the internal cavity field envelope amplitude, 
$k=2\pi/\lambda$ is the longitudinal wavenumber, $\Gamma'=cT/L$ is the 
amplitude decay rate from the cavity ($T$ is the amplitude 
transmission coefficient at each mirror, assumed equal, and $L$ is the 
cavity length), $\omega$ is the field angular frequency, 
$\Delta\omega'=\omega-\omega_{cav}$ is the detuning of the driving 
field from linear cavity resonance, and $n_2$ is the nonlinear index 
inside the cavity.  A time-independent version of this equation was 
also derived by Haelterman \textit{et al} from their modal theory 
\cite{haelterman1990}.  Equation \ref{dimcavityeqn} is slightly 
different from the version in Ref.  \cite{lugiato1987}, which has been 
rescaled to yield $\Gamma=1$; here we will employ a different rescaling
to easily accomodate the important limit $\Gamma\rightarrow0$.

Equation \ref{dimcavityeqn} is now made dimensionless by choosing an arbitrary 
distance scale $x_0$ and relating the time and field scales $t_0$ and
$|{\cal E}_0|$ to it using
$t_0 = k x_0^2 / c$ and
$|{\cal E}_0| = 1/{k x_0 \sqrt{|n_2|}}$.
After rescaling Eq. \ref{dimcavityeqn} takes on the dimensionless form
\begin{equation}
\dot\Psi = \frac{i}{2}\nabla^2\Psi+i\eta|\Psi|^2\Psi
+i\Delta\omega\Psi-\Gamma(\Psi-\Psi_{drive})\,,
\label{cavityeqn}
\end{equation}
where $\eta = +1 (-1)$ for self-focusing (-defocusing),
$\Psi={\cal E}/|{\cal E}_0|$, $\Delta\omega=\Delta\omega' t_{0}$, and
$\Gamma=\Gamma' t_{0}$.  The conditions of Eq. \ref{conditions} have
allowed us to use the mean field limit in deriving this equation and
also imply that $\Gamma\ll 1$.


\section*{Soliton Filtering and the Optical Whistle}

Equation \ref{cavityeqn} is a damped, driven version of the nonlinear
Schr\"{o}dinger equation.  In the limit
$\Gamma\rightarrow 0$ and for $\eta=+1$ it admits the family of
stationary soliton solutions \cite{chiao1993}
\begin{equation}
\begin{array}{rcl}
\Psi(X,T)&=&d^{-1}\,{\mathrm sech}(X/d)\\
\Delta\omega&=&-1/(2d^2)
\end{array}
\label{sechsolution}
\end{equation}
where $d$ is an arbitrary width and $X=x/x_0$, $T=t/t_0$
are dimensionless coordinates.  There are other, time-dependent
(``breather''), soliton solutions \cite{agrawal1989a} which are
unimportant for our present purposes.
Equation \ref{sechsolution} is also a solution in the damped case
($\Gamma\neq0$) if the driving field has the special form
\begin{equation}
\Psi_{drive,sech}(X,T) = \Psi(X,T)=d^{-1}\,{\mathrm sech}(X/d)\,.
\label{sechdrive}
\end{equation}
Whether this $\Gamma\neq0$ solution is a soliton in the mathematical 
sense is a subtle issue and not resolved here.  We will take the 
common practical approach and refer to this as a ``soliton solution'' 
in reference to its observed soliton-like robustness, keeping in mind 
the potential inappropriateness of this terminology.

A general qualitative feature of solitons is their 
robustness to small perturbations.  Hence we might expect that the 
Gaussian driving field
\begin{equation}
\Psi_{drive,Gaussian}(X,T) = d^{-1}\,{\mathrm exp}\left[-X^{2}/(1.699d)^{2}\right]\,.
\label{gaussiandrive}
\end{equation}
may lead to the soliton solution (\ref{sechsolution}) nearly as
efficiently as the sech drive (\ref{sechdrive}) when $\Gamma\ll 1$.
The constant 1.699 has been chosen to minimize the difference integral
\begin{equation}
\int_{-\infty}^{\infty}\left[\Psi_{drive,Gaussian}(X)-
\Psi_{drive,sech}(X)\right]^{2}\,dx\,.
\end{equation}
This possibility
was considered in a travelling wave geometry by Burak and Nasalski
\cite{burak1994}, who found that as much as 99.5\% of the incident
power is converted into a soliton.

In our present situation it is found that the folded path arising from 
the cavity configuration can modify this picture substantially.  
Generally, soliton forming behavior is observed when the driving field 
amplitude is matched to its width, as in Eq.  \ref{gaussiandrive}.  
Because the non-sech incident beam profile yields a sech beam in 
transmission, we call this process \textit{soliton filtering} in 
analogy with ordinary spatial filtering.  A larger driving amplitude 
leads to a ``mode mismatch'' since the soliton is trying to form with 
a smaller width than the input beam.  The optical whistle oscillation 
results from a competition between the soliton trying to shrink in 
size, and the driving field feeding in light mismatched in phase and 
spatial profile.

Considering again the Gaussian driving field of Eq.  
\ref{gaussiandrive}, recall that the choice of distance scale $x_0$ in 
deriving Eq. \ref{cavityeqn} is arbitrary, so we can choose it 
to be the waist size of the input beam without loss of generality.  
(In this case the time scale $t_0$ is the time of flight through the 
beam's Rayleigh range.)  Then we have
$\Psi_{drive}(X,T) = q\,\exp{(-X^2)}$ where $q$ is a real,
dimensionless driving amplitude, and our model is characterized by the
four real dimensionless quantities $\eta$, $\Gamma$, $\Delta\omega$, and
$q$.  We expect optimal coupling into the soliton when
\begin{equation}
\begin{array}{rcl}
q&\approx&1.699 \\
\Delta\omega&\approx&-(1.699)^{2}/2=-1.443\,. \\		
\end{array}
\label{goodparams}
\end{equation}
The former is
remarkably close to the optimal value $1.69$ in the travelling wave
case, found using the inverse scattering transform \cite{burak1994}.

\section*{Numerical Results}

Equation \ref{cavityeqn} is solved using a version of the popular 
split-step technique \cite{agrawal1989b}\cite{taha1984}, accurate 
through two orders in the time step size $h$.  The results presented 
below are calculated using $h=0.01$ and a spatial grid consisting of 
256 points separated by 0.156 dimensionless units.  These values are 
found to give accurate results for the cases of interest here; halving 
the grid spacings in space and time yields essentially identical 
results for a variety of test cases.

\subsection*{Oscillation}

\begin{figure}
\centerline{\psfig{figure=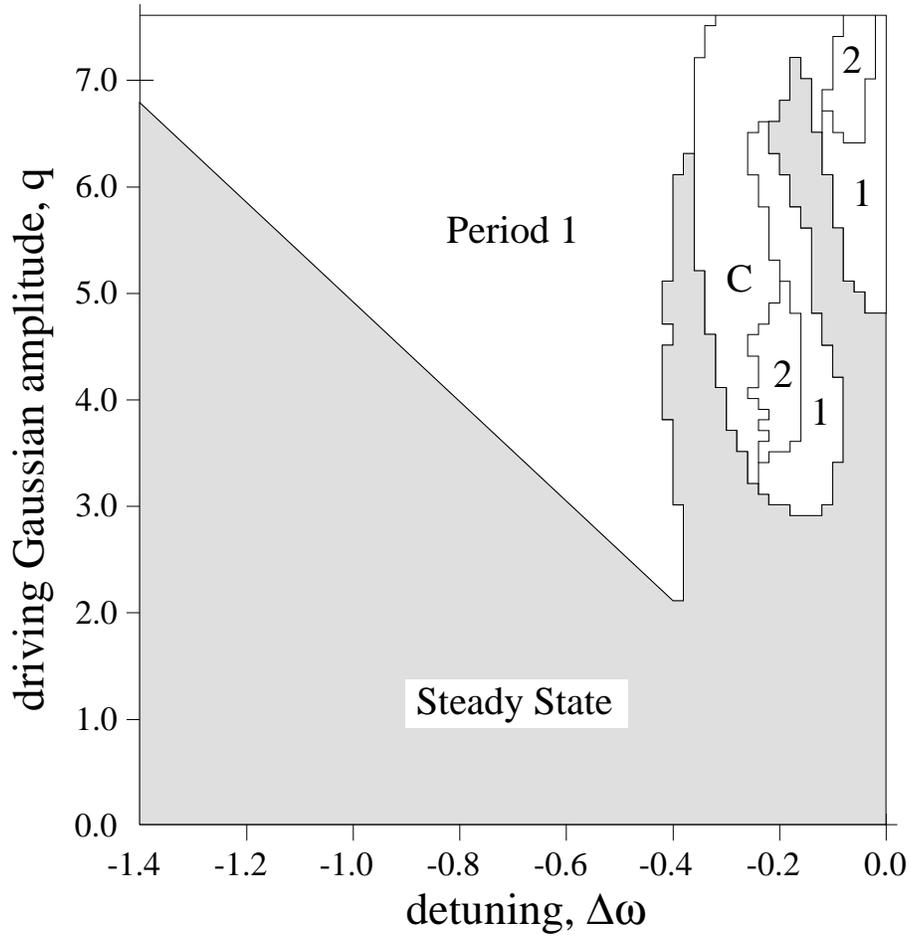,width=12cm}}
\caption{Asymptotic behavior of cavity field, for $\eta = +1$,
$\Gamma = 0.14$, and an initially empty cavity.  Labels indicate
regions of steady state, period 1 oscillation (1), period 2 
oscillation (2), and chaos (C).}
\label{paramspace}
\end{figure}

Figure \ref{paramspace} summarizes the asymptotic (long-time) cavity 
solutions when the Gaussian driving field is suddenly turned on at 
$t=0$.  We have here assumed an initially empty cavity, $\eta=+1$, and 
$\Gamma=0.14$.  The asymptotic cavity solutions are categorized into 
five basic types: (1) steady state solutions, (2) period 1 
oscillations, (3) period 2 oscillations, (4) long period oscillations, 
and (5) chaos.  The regions are seen to have complex boundaries, and 
points near the borders of the ``chaotic'' regions show particularly 
interesting behaviors: at the borders with steady state regions lie 
oscillations of very long period, and at the borders with the normal 
oscillating regions period doubling occurs.  Period 3 oscillations are 
also observed.

\begin{figure}
\centerline{\psfig{figure=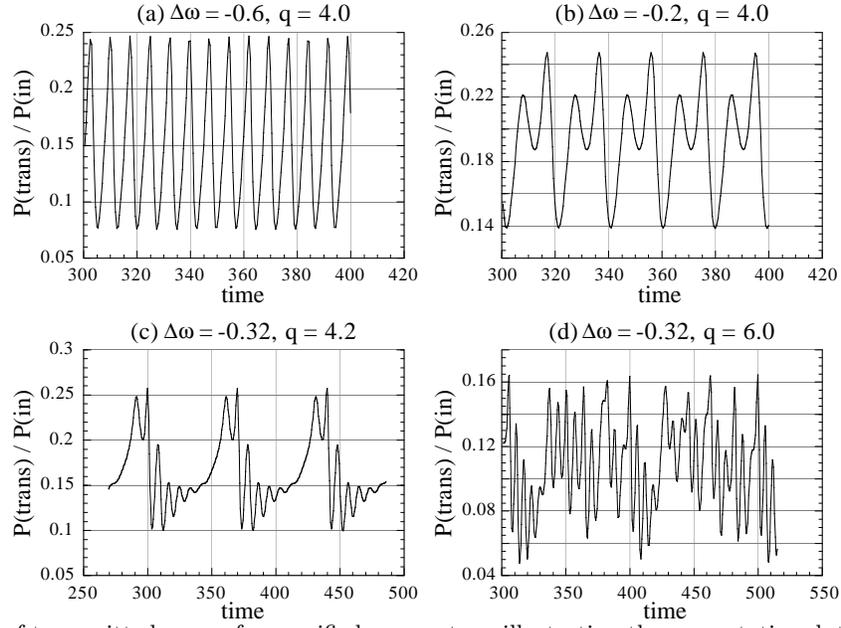,width=12cm}}
\caption{Time series of transmitted power for specified parameters,
illustrating the asymptotic solutions shown in Fig. \ref{paramspace}.
(a) period 1, (b) period 2, (c) long period, and (d) chaotic
oscillation.  $\eta=+1$, $\Gamma=0.14$.}
\label{timeseries}
\end{figure}

\begin{figure}
\centerline{\psfig{figure=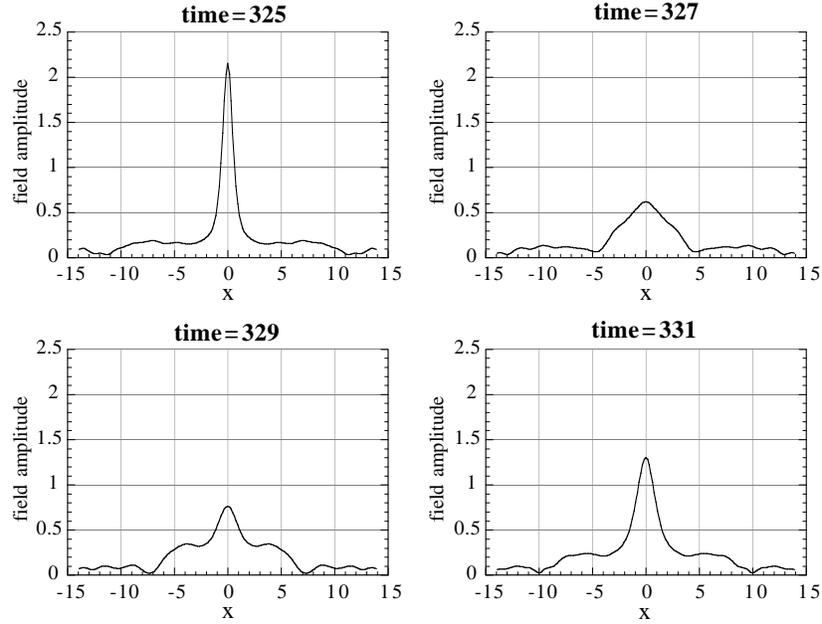,width=12cm}}
\caption{Cavity field amplitude profiles for the period 1 oscillation
shown in Figure \ref{timeseries}(a).  $\Delta\omega = -0.6$,
$q=4.0$,$\Gamma=0.14$.}
\label{snapshots}
\end{figure}

\begin{figure}
\centerline{\psfig{figure=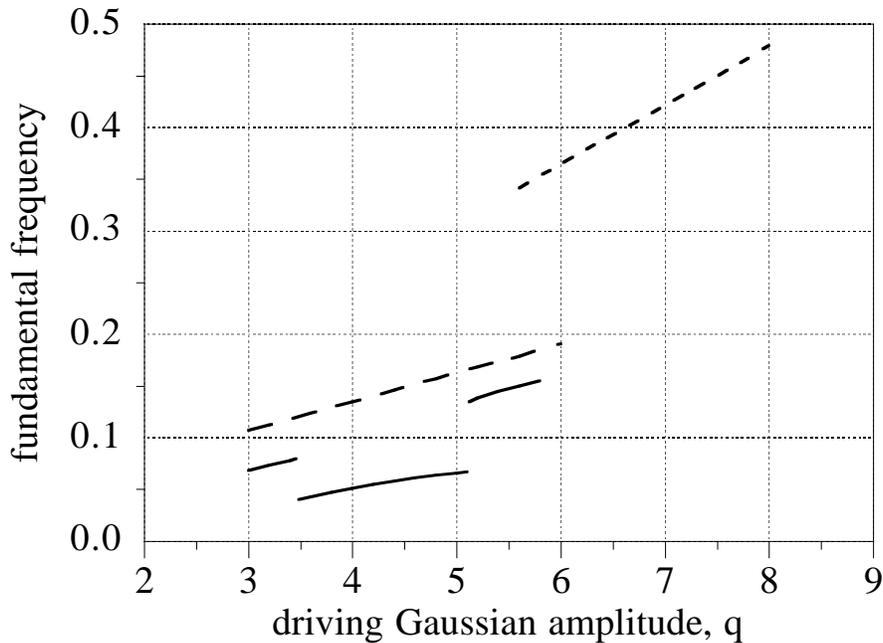,width=12cm}}
\caption{Fundamental frequency of oscillation as a function of
driving amplitude $q$.  The solid curve is for $\Gamma=0.14$,
$\Delta\omega=-0.2$ and shows one of the period-doubled
regions of Fig. \ref{paramspace}.  The long dashed curve is
for $\Gamma=0.14$, $\Delta\omega=-0.6$.  The short dashed curve
is for $\Gamma=0.3$, $\Delta\omega=-0.6$.}
\label{freqs}
\end{figure}

Time series of examples are shown in Figure \ref{timeseries}, where we 
plot the total power output for a variety of cases.  For the period 1 
example, snapshots of the cavity field amplitude are shown in Figure 
\ref{snapshots}.  The field profile narrows as the soliton forms, then 
is attenuated as the nonlinearity shifts the peak out of cavity 
resonance and power enters at the edges.  The fundamental oscillation 
frequencies for several cases are plotted in Figure \ref{freqs}.  It 
is not known what ranges of $\Gamma$ permit oscillations to occur.

\subsection*{Bistability}

\begin{figure}
\centerline{\psfig{figure=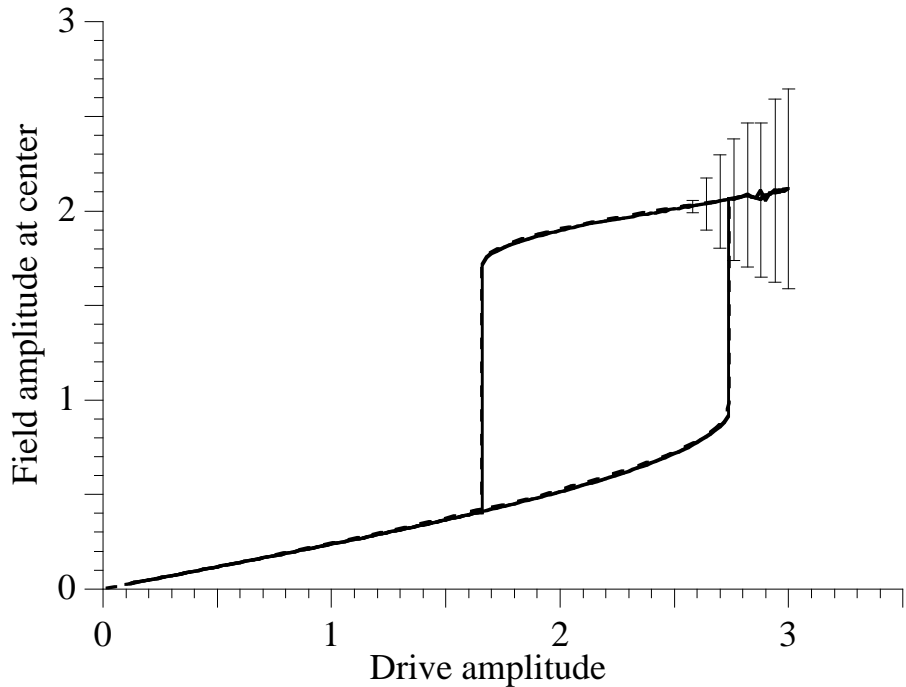,width=12cm}}
\caption{Comparison with steady state theory.  The dashed bistability
curve is from Fig. 3a of Vitrant \textit{et al} (1990)
and the solid curve
is from our time-dependent analysis.  Bars on the solid curve
show the oscillation amplitude.  $\Gamma=0.5$, $\Delta\omega=-1.5$.}
\label{comparison}
\end{figure}

An interesting issue is the effect the optical whistle has on 
dispersive optical bistability.  Figure \ref{comparison} compares a 
bistability curve from our time-dependent analysis with that given in 
Ref.  \cite{vitrant1990} which assumed a steady state field.  Our 
respective equations have been made dimensionless in slightly 
different manners; to correspond to the $X_{0}=1$, $\Delta=+3$ curve 
in their Figure 3a we have chosen $\Gamma=X_{0}^{2}/2=0.5$, 
$\Delta\omega=-\Gamma\Delta=-1.5$ and have related their field 
amplitude $f$ to ours using $\Psi=\sqrt{\Gamma}f$.  Part of the upper 
branch of the bistability curve is seen to be unstable due to the 
whistle oscillation, and the oscillation amplitude is indicated by the 
bars in Fig.  \ref{comparison}.  In general we observe that where 
bistability occurs ($\Delta\omega<-0.19$ for $\Gamma=0.14$) 
the whistle oscillation occurs only in the upper branch of the 
bistability curve.  However, the oscillation can also occur when there is 
no bistability, as is evident from Fig.  \ref{paramspace}.

\begin{figure}
\centerline{\psfig{figure=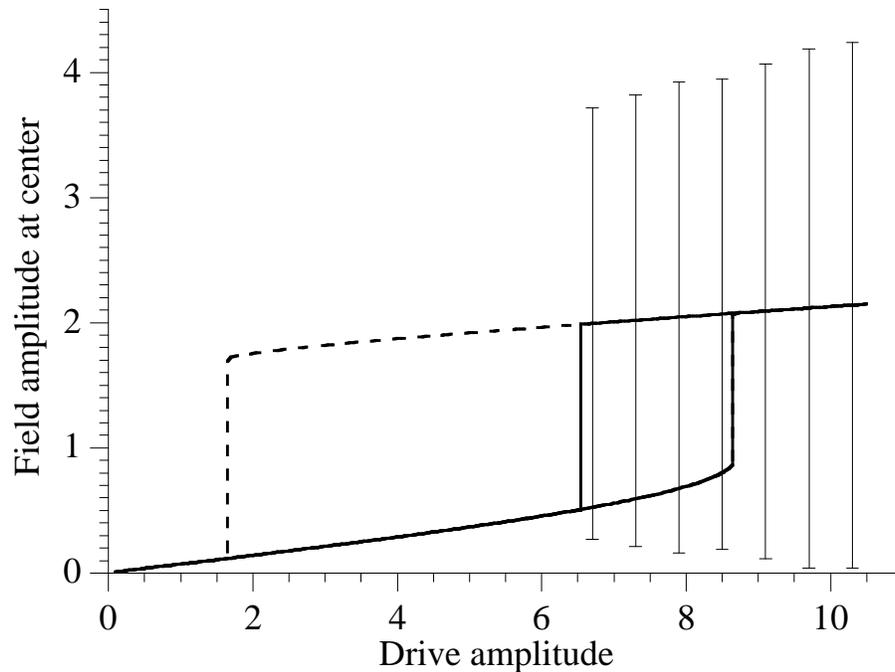,width=12cm}}
\caption{Comparison of steady state (dashed) and time-dependent 
(solid, with oscillation amplitude indicated by bars) analyses.  In
this case the whistle instability changes the switch-off point in
the bistability curve.  $\Gamma=0.14$, $\Delta\omega=-1.443$.}
\label{disruption}
\end{figure}

In other cases the optical whistle modifies bistability curves more
dramatically.  Figure \ref{disruption} shows steady state and
time-dependent bistability curves for $\Gamma=0.14$, 
$\Delta\omega=-1.443$.  In this case the whistle oscillation has moved
the switch-off point substantially, a common occurrence.
In contrast, the switch-on point has not been seen to
change from its steady state value in our analysis.

\subsection*{Soliton Formation}

\begin{figure}
\centerline{\psfig{figure=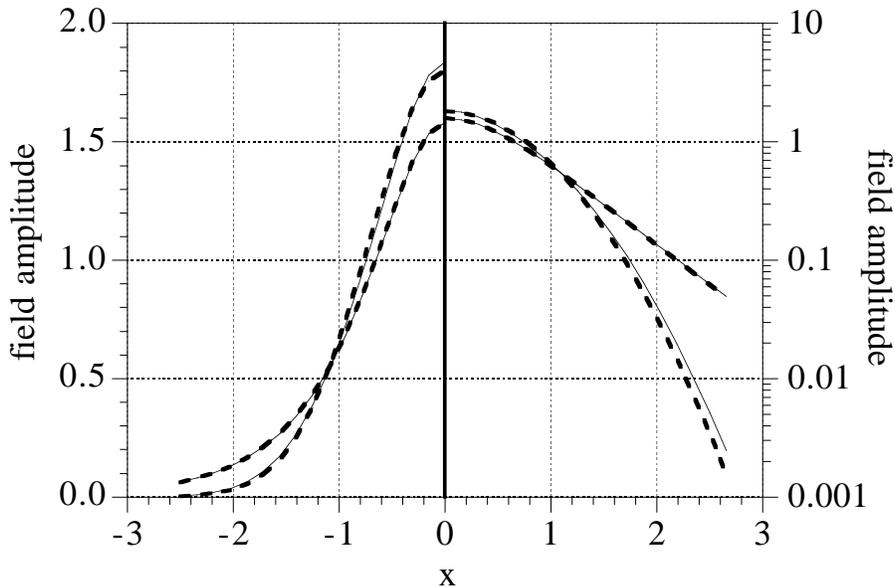,width=12cm}}
\caption{Steady state field solution with Gaussian drive,
plotted on linear and logarithmic scales, in the upper transmission
branch.  The dashed curves are the driving
function and $1.58\,\mathrm{sech}(1.575x)$; overlapping solid curves are solutions
with $\Gamma=10$ and $\Gamma=0.1$, respectively.  The latter solution is
very close to the soliton form of Eq. \ref{sechsolution}
despite being driven by a Gaussian beam, providing evidence
for the stability of the sech solution when $\Gamma\ll1$.  $q=1.8$,
$\Delta\omega=-1.2$.}
\label{profiles}
\end{figure}

We argued above that the soliton solution of Eq.  \ref{sechsolution} 
would be stable when $\Gamma\ll 1$, in the sense that a sech-profiled 
cavity field would result when a Gaussian driving beam is applied.  
This is numerically demonstrated in Figure \ref{profiles} where steady 
state field solutions are presented for the cases $\Gamma=0.1$ and 
$\Gamma=10$.  The former solution is very close to the sech form of 
Eq.  \ref{sechsolution}, and the latter is close to the Gaussian 
driving field.  Although the existence of solitons in a damped cavity 
is an unresolved technical question, the sech solutions are seen to be 
quite stable when $\Gamma\ll 1$.

\begin{figure}
\centerline{\psfig{figure=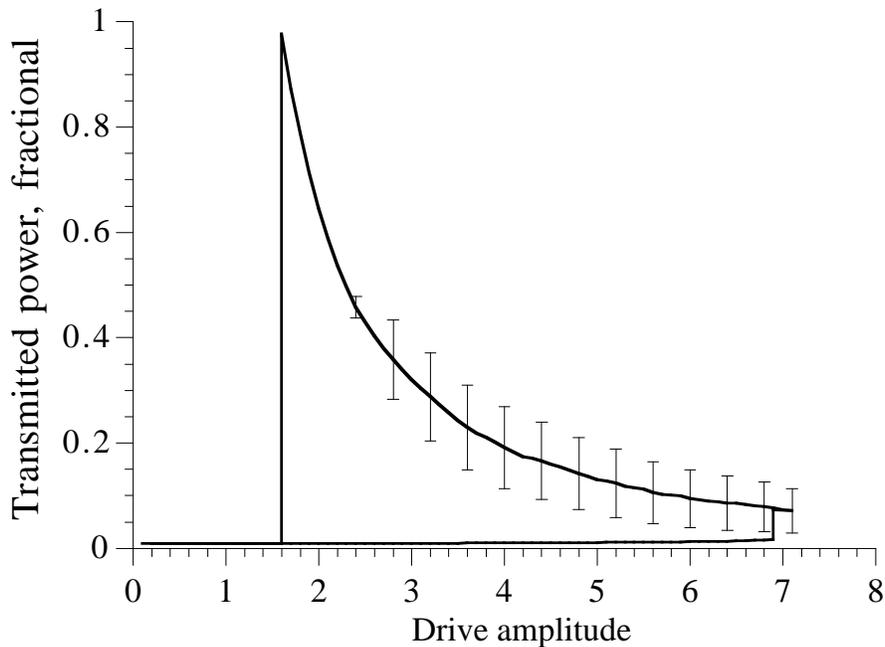,width=12cm}}
\caption{Bistability curve showing nearly complete (97.6\%) coupling
into the transmitted sech beam.  The expected optimal case,
Eq. \ref{goodparams}, is disrupted by the whistle as shown in
Fig. \ref{disruption}.  $\Delta\omega=-1.2$, $\Gamma=0.14$.}
\label{goodcoupling}
\end{figure}

\begin{figure}
\centerline{\psfig{figure=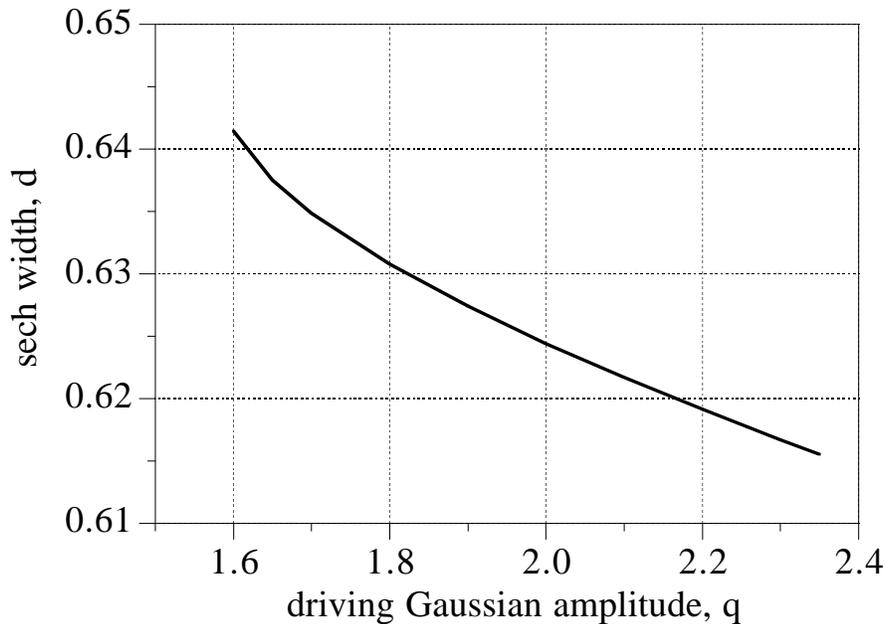,width=12cm}}
\caption{Width $d$ of soliton solution in upper branch of
Fig. \ref{goodcoupling} as a function of driving amplitude
$q$.  As $q$ increases the soliton narrows in width and
begins oscillating at $q\approx 2.4$.
$\Delta\omega=-1.2$, $\Gamma=0.14$.}
\label{width}
\end{figure}

In the travelling wave case soliton formation with Gaussian driving 
beams can be very efficient; as much as 99.5\% of the input power can 
be transformed into a soliton \cite{burak1994}.  In a cavity we might 
expect this to be an upper bound since imperfect interference (and 
reflection of the incident beam) would seem to present an additional 
loss of efficiency.
In Eq.  \ref{goodparams} we made a simple 
estimate of the optimal coupling parameters.  However, Fig.  
\ref{disruption} shows that for $\Delta\omega=-1.443$, $\Gamma=0.14$ 
the whistle disrupts the bistability curve's upper branch near 
$q=1.699$.  A nearby case ($\Delta\omega=-1.2$, $\Gamma=0.14$) without 
the disruption is shown in Fig.  \ref{goodcoupling}.  At the 
switch-off point $q\approx1.6$ the coupling efficiency into the 
transmitted beam is 97.6\%, still quite high.  Because $\Gamma=0.14\ll1$
the transmitted field is found to be quite close to the sech form 
of Eq.  \ref{sechsolution}; when $\Gamma>1$ the transmitted field is 
more Gaussian in profile.  Figure \ref{width} shows the steady state
soliton width in the upper branch of Fig. \ref{goodcoupling}, and the
decrease in width as the oscillation point is reached.
For the present work we have not attempted 
a thorough study to optimize soliton coupling efficiency but 
wish to point out that fairly high efficiencies are possible in 
cavities.

\section*{Discussion}

A variety of instability phenomena similar to the optical whistle have 
been reported in the literature \cite{reinisch1994}.  These can be 
broadly categorized into multi-mode and single-mode (mean field) 
effects.  In the former case the plane wave oscillation phenomenon has 
been called ``self-pulsing'' (it occurs in both the context of 
absorptive \cite{bonifacio1979} and dispersive 
\cite{ikeda1979}\cite{lugiato1980} bistability).  The instability here 
is essentially due to the interplay of two different timescales, 
nonlinearity and cavity feedback, and gives rise to an oscillation 
period which is a multiple of the cavity round-trip time.  Slightly 
later the single-mode case was also shown to yield oscillations.  
Ikeda and Akimoto \cite{ikeda1982} first demonstrated this within a 
plane wave, purely dispersive Kerr model.  Lugiato \textit{et al} 
\cite{lugiato1982} constructed a more realistic optical Bloch model 
and found oscillations (and chaos) in that case as well, again at the 
plane wave level of description.

Our work is more closely related to the transverse instability 
discovered by McLaughlin \textit{et al} \cite{mclaughlin1983} and 
discussed in the mean field limit by Lugiato and Lefever 
\cite{lugiato1987}.  In these studies, stationary transverse structure 
in a broad illuminating beam arises from a modulational instability.  
McLaughlin \textit{et al} \cite{mclaughlin1985} showed in 1985 that 
the transverse features can oscillate in time.  In all cases, more 
recent numerical work has shown that these transverse perturbations 
can grow to become soliton chains 
\cite{haelterman1992}\cite{mcdonald1990} which may or may not be 
stationary in time.  Lastly, for a plane wave driving field of 
variable intensity a series of bifurcations similar to the ones seen 
here are observed to occur in the mean field case, eventually leading 
to spatio-temporal chaos \cite{nozaki1986}.  Our present work is 
different from these earlier results because here the most unstable 
transverse wavelength is comparable to the driving field width; the 
instability consists of a broad spectrum of transverse wavenumbers and 
it is somewhat more intuitive to view the process as a mode mismatch 
in the soliton formation process rather than as a sinusoidal 
modulation in a very broad beam growing into a (perhaps) time-varying 
soliton chain.

There are several possible experimental configurations in which the
optical whistle might be observed. One concrete realization is a fiber-loop
geometry \cite{mitschke1997}, in which Gaussian pulses from
a laser are injected into a (nonlinear) fiber loop and the 
copropagating longitudinal coordinate $z-vt$ plays the role of the
tranverse dimension $x$ in our analysis.  Another possibility is a 
cylindrical nonlinear cavity with strong single-moding in the $y$ and $z$ 
dimensions \cite{deutsch1992}.

The optical whistle phenomenon presented here is believed to exist 
beyond the Kerr nonlinear, one dimensional model, whenever 
one is in the \textit{soliton forming limit} of Eq.  \ref{conditions}.  
We conjecture this on the basis of our general picture that the 
whistle is the result of a mode mismatch between the generated soliton 
and driving field.  The study of related phenomena in other systems 
would be an interesting topic of further research.

\section*{Acknowledgements}
The authors would like to thank Morgan W. Mitchell and Prof.  Ewan M. 
Wright for very helpful discussions.  We are also appreciative of the 
referee's insightful comments and suggestions.

\end{document}